\documentclass[12pt]{iopart}

\usepackage[utf8]{inputenc}
\usepackage{a4wide}

\expandafter\let\csname equation*\endcsname\relax

\expandafter\let\csname endequation*\endcsname\relax

\usepackage{amsmath}
\usepackage{amssymb,amsthm,mathrsfs}
\usepackage{dsfont}
\usepackage{graphicx}

\usepackage{verbatim}
\usepackage{epstopdf}
\usepackage{xcolor}

\newcommand{\ket}[1]{|#1\rangle}
\newcommand{\braket}[2]{\langle#1|#2\rangle}
\newcommand{\ketbra}[2]{|#1\rangle\langle #2|}
\newcommand{\braketA}[3]{\langle#1|#2|#3\rangle}

\begin{document}

\title{Quantum walk state transfer on a hypercube}

\author{Martin \v{S}tefa\v{n}\'{a}k$^*$ and Stanislav Skoupý}

\address{Department of Physics, Faculty of Nuclear Sciences and Physical Engineering, Czech Technical University in Prague, B\v rehov\'a 7, 115 19 Praha 1 - Star\'e M\v esto, Czech Republic \\
$^*$ corresponding author}
\ead{martin.stefanak@fjfi.cvut.cz; stanislav.skoupy@fjfi.cvut.cz}
\vspace{10pt}
\date{\today}

\begin{abstract}
We investigate state transfer on a hypercube by means of a quantum walk where the sender and the receiver vertices are marked by a weighted loops. First, we analyze search for a single marked vertex, which can be used for state transfer between arbitrary vertices by switching the weighted loop from the sender to the receiver after one run-time. Next, state transfer between antipodal vertices is considered. We show that one can tune the weight of the loop to achieve state transfer with high fidelity in shorter run-time in comparison to the state transfer with a switch. Finally, we investigate state transfer between vertices of arbitrary distance. It is shown that when the distance between the sender and the receiver is at least 2, the results derived for the antipodes are well applicable. If the sender and the receiver are direct neighbours the evolution follows a slightly different course. Nevertheless, state transfer with high fidelity is achieved in the same run-time. 
\end{abstract}

\maketitle

\section{Introduction}

One of the fruitful application of quantum walks \cite{Aharonov1993,Meyer1996,Farhi1998} is the task of state transfer \cite{bose2003} between two vertices of a graph. Consider the particle initially localized on the sender vertex. Our goal is to transfer it with high probability to the receiver vertex by a quantum walk evolution. There are two basic approaches to this problem. If we have control over all sites we can globally design the dynamics such that the walker is transferred between a selected pair of vertices at a certain time. For the continuous time evolution, where the dynamics is governed by the Schroedinger equation with a given hamiltonian, the problem was investigated to large detail in  the context of spin chains \cite{christandl_perfect_2004,christandl_perfect_2005,plenio_high_2005,bose_2007,gualdi_perfect_2008,kay_perfect_2010}, discrete quantum networks of various topologies \cite{kostak_perfect_2007,nikolopoulos_analysis_2012,hoskovec_decoupling_2014,frydrych_selective_2015,hoskovec_dynamical_2022} and continuous time quantum walks \cite{kendon2011,godsil_state_2012,godsil_state_2020,coutinho_perfect_2015,chen_pair_2020}. In the discrete time case this approach was studied on various graphs such as circle \cite{kurzynski2011,yalcinkaya2015}, regular graphs \cite{shang2018} or more general networks \cite{chen2019}. Recently, discrete time quantum walks with evolution operators defined by a product of two reflections \cite{zhan_quantum_2021} were also applied for state transfer. The paper \cite{kubota_perfect_2022} focused on walks with the Grover coin, while \cite{chan:2021}, \cite{guo:2022} investigated a more general setting of coins constructed from weighted tail-arc incidence matrices, finding several classes of graphs supporting state transfer.

The second approach to state transfer utilizes quantum spatial search based on quantum walks \cite{aaronson_quantum_2003}. Over the years this was a very fruitful field of research, providing quadratic speed-up over classical search e.g. on hypercube \cite{Shenvi2003,Potocek2009,hein2009:search}, lattices \cite{Childs2004,childs_2004b,Ambainis2005} or symmetric graphs \cite{reitzner2009}, \cite{chakraborty2016}. Later, it was shown that the search utilizing continuous time quantum walk is optimal for almost all graphs \cite{chakraborty2016}. More recently, quadratic speed-up over a classical random walk search for any number of marked vertices was achieved in both discrete time \cite{ambainis_quadratic_2020} and continuous time \cite{apers_quadratic_2022} quantum walk algorithms. In this approach 
the dynamics at the sender and receiver vertices is altered locally. In quantum walk search we evolve the system from the equal weight superposition to a state localized on the marked vertex. For state transfer the walk starts localized on the sender vertex and evolves through the equal weight superposition to a state localized at the receiver vertex. One possibility is to mark both vertices that want to communicate at the same time. In the discrete time case this method was proposed for state transfer on lattices \cite{hein2009} and further analyzed on various types of finite graphs, e. g. on cycles and their variants \cite{kendon2011,barr2014}, star and complete graph with loops \cite{stefanak2016}, complete bipartite graph \cite{stefanak2017}, circulant graphs \cite{zhan2019} or butterfly network \cite{cao2019}. In the continuous time evolution this approach was investigated on Erdos-Renyi graphs \cite{chakraborty2016}. Nevertheless, running the search algorithm with two marked vertices does not necessarily guarantee state transfer with fidelity close to 1. However, if the search success probability is high, one can use the switch approach \cite{skoupy:2022,Santos_2022}. Here we initially mark only the sender vertex and evolve the search algorithm for one period to reach a state close to the equal weight superposition, then switch the marking from the sender to the receiver vertex and evolve the search algorithm for one more period. 

In comparison with the first method the modification of search for state transfer has the advantage that in principle any two pairs of vertices can establish communication without the need to adjust the dynamics on the rest of the graph. They don't even have to know each others location, only some global property of the graph e.g. number of vertices which influences the run-time.  

In the present paper we consider state transfer on a hypercube by a discrete time quantum walk where the sender and the receiver vertices are marked with a weighted loop. First, we investigate the search for a single marked vertex, which is then directly utilized to perform state transfer with a switch. We follow the approach of \cite{Shenvi2003} and reduce the problem to a walk on a finite line with position dependent coins. The difference to the original search algorithm is that the resulting approximate invariant subspace is three-dimensional, as the loop effectively adds an eigenvector corresponding to an eigenvalue 1. The optimal weight resulting in high success probability is determined to be $n/2^n$, which is in accordance with previous studies \cite{wong2015,wong2018,rhodes2019,rhodes2020,chiang2020,hoyer2020}.

Next, we focus on the state transfer between marked vertices which are antipodal. In such a case, the problem still can be reduced to a finite line, however, the approximate invariant subspace has dimension 5. Apart from an eigenvalue 1 there are two pairs of complex conjugated eigenvalues. We show that state transfer with high fidelity can be achieved by tuning the weight such that the phases of the eigenvalues become harmonic. The resulting quantum walk performs state transfer faster than the approach with the switch.

Finally, we consider state transfer between marked vertices at arbitrary distance. Here we mostly rely on numerical simulations, nevertheless, we determine an exact 1-eigenvector of the evolution operator relevant for the dynamics of state transfer. When the sender and the receiver vertices are not connected by an edge the evolution of the system is very close to the case of the antipodal vertices. The case of state transfer between direct neighbours is a bit distinct. We show that the system evolves approximately in a three-dimensional subspace. Nevertheless, state transfer with high fidelity is achieved for the same parameters, i.e. weight of the loop and the number of steps, as before. 

The rest of the paper is organized as follows: we set the notation and investigate the search for a single marked vertex and state transfer with switch in Section \ref{sec:search}. In Section \ref{sec:antipode} we analyze in detail the state transfer between antipodal vertices. The investigation is extended to vertices of arbitrary distance in Section \ref{sec:arbitrary}. Finally, we conclude and present an outlook in Section \ref{sec:concl}.

\section{Search on a hypercube with a weighted loop and state transfer with a switch}
\label{sec:search}

Let us begin with the description of an unperturbed Grover walk on an $n$-dimensional hypercube. We adopt the notation from the seminal paper \cite{Shenvi2003} on the quantum walk search on a hypercube. The graph has $2^n$ vertices which are labeled by $n$-bit strings $\vec{x} = x_1\ldots x_n$, where $x_i =0,1$. To each vertex we assign a basis vector $\ket{\vec{x}}$ of the position space
$$
{\mathcal H}_P = {\rm Span}\left\{\ket{\vec{x}}|\vec{x} = x_1\ldots x_n,\ x_i =0,1\right\} .
$$
Each vertex of the hypercube has degree $n$, the neighbouring vertices have Hamming distance 1, i.e. they differ in a single bit. Since hypercube is regular, we can introduce the coin space 
$$
{\mathcal H}_C = {\rm Span} \left\{\ket{d}|d=1,\ldots n\right\},
$$
in the same way at every vertex. The Hilbert space of the quantum walk on the hypercube is then
$$
{\mathcal H } = {\mathcal H}_P\otimes {\mathcal H}_C.
$$
The unitary evolution operator of the walk is a product of the shift and the coin operator
$$
U = S\cdot C.
$$
The shift is given by
$$
S = \sum_{d=1}^n \sum_{\vec{x}} \ketbra{\vec{x}\oplus\vec{e}_d,d}{\vec{x},d},
$$
where $\vec{e}_d$ is a basis vector in the direction $d$, i.e. $\vec{e}_d = e^d_1\ldots e^d_n$ with $e^d_j = \delta_{j,d}$, and $\oplus$ denotes addition modulo 2. For the unperturbed walk the coin operator is homogeneous across the graph, i.e.
$$
C = I_P \otimes G
$$
where $I_P$ is the identity in the position space and $G$ is the Grover diffusion operator
$$
G = 2\ketbra{s_C}{s_C} - I_C, \quad \ket{s_C} = \frac{1}{\sqrt{n}} \sum_{d=1}^n \ket{d}. 
$$

Turning to the quantum search, we have to perturb the walk. We attach a loop to the marked vertex (but only at the marked vertex) and consider a different coin. The graph is no longer regular, since the marked vertex which can be labeled without loss of generality as $\vec{0}$, has a degree $n+1$. The Hilbert space of the perturbed walk is now given by a direct sum
$$
{\mathcal H}' = \bigoplus_{\vec{x}} {\mathcal H}_{\vec{x}}
$$
of local Hilbert spaces
\begin{align*}
\vec{x}\neq \vec{0} : \quad & \quad  {\mathcal H}_{\vec{x}} = {\rm Span}\left\{\ket{\vec{x},d}|d=1,\ldots n\right\}  , \\
\vec{x} = \vec{0} : \quad & \quad  {\mathcal H}_{\vec{0}} = {\rm Span}\left\{\ket{\vec{0},d}|d=0,1,\ldots n\right\} .
\end{align*}
The evolution operator is modified into
$$
U' = S' \cdot C',
$$
where the shift is given by
$$
S' = \sum_{d=1}^n \sum_{\vec{x}} \ketbra{\vec{x}\oplus\vec{e}_d,d}{\vec{x},d} + \ketbra{\vec{0},0}{\vec{0},0} . 
$$
For the coin we keep the Grover diffusion operator $G$ on the non-marked vertices and alter it at the marked vertex $\vec{0}$ to $-G'$
$$
C' = (I_P - \ketbra{\vec{0}}{\vec{0}}) \otimes G - \ketbra{\vec{0}}{\vec{0}}\otimes G' . 
$$
Here $G'$ denotes the modified Grover coin with a loop of weight $l$
$$
G' = 2\ketbra{s_l}{s_l} - I_0, \quad \ket{s_l} = \frac{1}{\sqrt{n+l}}\left(\sqrt{l}\ket{0} + \sum_{d=1}^n \ket{d} \right) .
$$
For the initial state of the search, we consider the equal weight superposition of all basis states except for the state of the loop, i.e.
$$
\ket{\psi_0} = \frac{1}{\sqrt{n 2^n}} \sum_{d=1}^n \sum_{\vec{x}} \ket{\vec{x},d} . 
$$
In the same way as in \cite{Shenvi2003}, the problem can be reduced to the walk on a finite line with a non-homogeneous coin. Define the following $2n+1$ orthonormal basis states
\begin{align*}
    \ket{x,R} & = \frac{1}{\sqrt{(n-x){n\choose x}}} \sum_{|\vec{x}| = x}\sum_{x_d = 0} \ket{\vec{x},d}, \quad x = 0,\ldots n-1 , \\
    \ket{x,L} & = \frac{1}{\sqrt{x{n\choose x}}} \sum_{|\vec{x}| = x}\sum\limits_{x_d=1} \ket{\vec{x},d}, \quad x = 1,\ldots n, \\
    \ket{0,\circlearrowleft} & = \ket{\vec{0},0} .
\end{align*}
The unperturbed shift operator acts in this basis as
$$
S = \sum_{x=0}^{n-1} \left(\ketbra{x,R}{x+1,L} + \ketbra{x+1,L}{x,R}\right) $$
while for the perturbed shift we have to add the loop at 0
$$
S' = S + \ketbra{0,\circlearrowleft}{0,\circlearrowleft} .    $$
The coins $C$ and $C'$ acts locally on each position $x$, and their actions differ only at the marked vertex 0
\begin{align*}
C & = \sum_{x=0}^n \ketbra{x}{x}\otimes C_x, \\
C' & = \ketbra{0}{0}\otimes C_0' +   \sum_{x=1}^n \ketbra{x}{x}\otimes C_x .
\end{align*}
The local coin operators for $x\neq 0, n$ are in the $\left\{\ket{R},\ket{L}\right\}$ basis given by
$$
C_x = \begin{pmatrix}
\cos\theta_x & \sin\theta_x \\
\sin\theta_x & -\cos\theta_x
\end{pmatrix},  
$$
where the angle $\theta_x$ is determined by 
$$
\quad \cos\theta_x = 1 - \frac{2x}{n}, \quad \sin\theta_x = \frac{2}{n}\sqrt{x(n-x)}
$$
For $x=n$ the coin space is one-dimensional and the coin reduces to $C_n = 1$, and the same holds for $x=0$ in the unpertubed case. For the perturbed case, in contrast to \cite{Shenvi2003}, the coin space at the marked vertex $x=0$ is two-dimensional, and in the basis $\left\{\ket{R},\ket{\circlearrowleft}\right\}$ the local coin operator reads
$$
C_0' = \begin{pmatrix}
-\frac{n-l}{n+l} & -\frac{2\sqrt{n l}}{n+l} \\
-\frac{2\sqrt{n l}}{n+l} & \frac{n-l}{n+l}
\end{pmatrix} .
$$
The initial state of the search is in the basis $\left\{\ket{x,R},\ket{x,L},\ket{0,\circlearrowleft}\right\}$ expressed as
\begin{align*}
\ket{\psi_0} = & \ \frac{1}{2^\frac{n}{2}}(\ket{0,R} + \ket{n,L}) + \frac{1}{2^\frac{n}{2}}\sum_{x=1}^{n-1} \left(\sqrt{{n-1\choose x-1}}\ket{x,L} + \sqrt{{n-1\choose x}}\ket{x,R} \right) .    
\end{align*}

Let us show that the search algorithm evolves approximately in a three-dimensional subspace spanned by $\{\ket{\psi_0},\ket{0\circlearrowleft},\ket{\psi_1}\}$,
where\footnote{Similar state was employed in the analysis of \cite{Shenvi2003}, but here we modify it slightly - we omit the last term in the summation, which simplifies the calculations for the state transfer.}
\begin{align*}
    \ket{\psi_1} & = \frac{1}{c} \sum_{x=0}^{n/2-2} \frac{1}{\sqrt{2{n-1\choose x}}} \left(\ket{x,R} -  \ket{x+1,L}\right), \quad c  = \sqrt{\sum_{x=0}^{n/2-2} \frac{1}{{n-1\choose x}}}.
\end{align*}
Note that with the estimate
$$
\sum_{x=0}^{n/2-2} \frac{1}{{n-1\choose x}} \approx 1 + \frac{1}{n-1},
$$
we can approximate the normalization of $\ket{\psi_1}$ by
\begin{equation}
\label{norm:approx}
\frac{1}{c} \approx \sqrt{1-\frac{1}{n}} \approx 1 - \frac{1}{2n}.    
\end{equation}
First, by direct calculation one can show that
$$
\ket{\alpha_1} = \sqrt{\frac{l 2^n}{n+l 2^n}}\ket{\psi_0} - \sqrt{\frac{n}{n+l 2^n}}\ket{0,\circlearrowleft},
$$
is an exact eigenvector of $U'$ with eigenvalue 1. Second, for small values of the weight $l$ the states 
\begin{align*}
       \ket{\alpha_2} & = \sqrt{\frac{n}{n+l 2^n}}\ket{\psi_0} + \sqrt{\frac{l 2^n}{n+l 2^n}}\ket{0,\circlearrowleft}, \\
    \ket{\alpha_3} & = \ket{\psi_1} ,
\end{align*}
are approximate eigenvectors. We find that the action of $U'$ on $\ket{\alpha_{2,3}}$ is given by
\begin{align*}
 U' \ket{\alpha_2} = & \ \ket{\alpha_2} - \frac{2}{n+l}\sqrt{l+\frac{n}{2^n}}\left(\sqrt{l} \ket{0,\circlearrowleft} + \sqrt{n}\ket{1,L} \right), \\
U'\ket{\alpha_3} =  & \ \ket{\alpha_3} - \frac{\sqrt{2l}}{c(n+l)} \left( \sqrt{n}\ket{0,\circlearrowleft} - \sqrt{l}\ket{1,L}\right) - \\
& - \frac{2}{c\sqrt{2 {n-1\choose \lfloor \frac{n}{2}\rfloor-2}}} \left( \frac{\lfloor\frac{n}{2}\rfloor-1}{n}\ket{\lfloor \frac{n}{2}\rfloor-2,R} + \frac{\sqrt{(\lfloor\frac{n}{2}\rfloor-1)(n-\lfloor\frac{n}{2}\rfloor+1)}}{n}\ket{\lfloor\frac{n}{2}\rfloor,L} \right) ,
\end{align*}
from which we determine the matrix elements
\begin{align*}
\braketA{\alpha_2}{U'}{\alpha_2}  & = 1 - 2\frac{l+\frac{n}{2^n}}{n+l} , \\
\braketA{\alpha_3}{U'}{\alpha_3} & = 1 -  \frac{l}{c^2(n+l)} - \frac{\lfloor\frac{n}{2}\rfloor-1}{c^2 n {n-1\choose \lfloor\frac{n}{2}\rfloor-2}},\\
\braketA{\alpha_3}{U'}{\alpha_2}  & = - \braketA{\alpha_2}{U'}{\alpha_3} = \frac{1}{c}\frac{\sqrt{2n(l+\frac{n}{ 2^{n}})}}{n+l} . 
\end{align*}
In the basis $\left\{\ket{\alpha_1},\ket{\alpha_2},\ket{\alpha_3}\right\}$ the evolution operator of the search $U'$ can be approximated by an effective matrix
$$
U'_{ef} = \begin{pmatrix}
1 & 0 & 0 \\
0 & \cos{\omega} & \sin\omega \\
0 & -\sin\omega & \cos\omega
\end{pmatrix}
$$
with the angle $\omega$ given by
$$
\omega \approx \sin\omega = \braketA{\alpha_3}{U'}{\alpha_2} \approx \frac{(2n-1)\sqrt{l+\frac{n}{ 2^{n}}}}{\sqrt{2n}(l+n)} ,
$$
where we have used the approximation of the normalization (\ref{norm:approx}). Hence, the evolution keeps the state $\ket{\alpha_1}$ unchanged and acts as an approximate rotation in the $\ket{\alpha_2}$, $\ket{\alpha_3}$ plane by an angle $\omega$.

Let us now estimate the success probability for a given $l$. In terms of $\ket{\alpha_j}$, the initial state and the target state (loop at the marked vertex) are given by
\begin{align*}
\ket{\psi_0} & = \sqrt{\frac{l 2^n}{n+l 2^n}}\ket{\alpha_1} + \sqrt{\frac{n}{n+l 2^n}}\ket{\alpha_2}, \\
\ket{0,\circlearrowleft} & = - \sqrt{\frac{n}{n+l 2^n}}\ket{\alpha_1} + \sqrt{\frac{l 2^n}{n+l 2^n}}\ket{\alpha_2}.
\end{align*}
Maximal success probability is achieved if we rotate from $\ket{\alpha_2}$ to $-\ket{\alpha_2}$, i.e. if we reach a state
$$
\ket{\psi(T)} \approx \sqrt{\frac{l 2^n}{n+l 2^n}}\ket{\alpha_1} - \sqrt{\frac{n}{n+l 2^n}}\ket{\alpha_2}.
$$
In this state, we find the walker in the loop with probability
\begin{equation}
\label{ssl}
    P(l) = |\braket{0,\circlearrowleft}{\psi(T)}|^2 \approx \frac{4n l2^n}{(n+l 2^n)^2}.
\end{equation}
This function has a maximum equal to one for $l=\frac{n}{2^n}$, confirming that the optimal weight can be chosen as $l=\frac{d}{N}$ for a $d-$regular graph even if we consider the loop only at the marked vertex. Note that for $l=\frac{n}{2^n}$ the vectors $\ket{\alpha_{1,2}}$ simplify into
\begin{align*}
    \ket{\alpha_1} & = \frac{1}{\sqrt{2}}\left(\ket{\psi_0} - \ket{0,\circlearrowleft}\right), \\
    \ket{\alpha_2} & = \frac{1}{\sqrt{2}}\left(\ket{\psi_0} + \ket{0,\circlearrowleft}\right) .
\end{align*}

\begin{figure}
    \centering
    \includegraphics[width=0.6\textwidth]{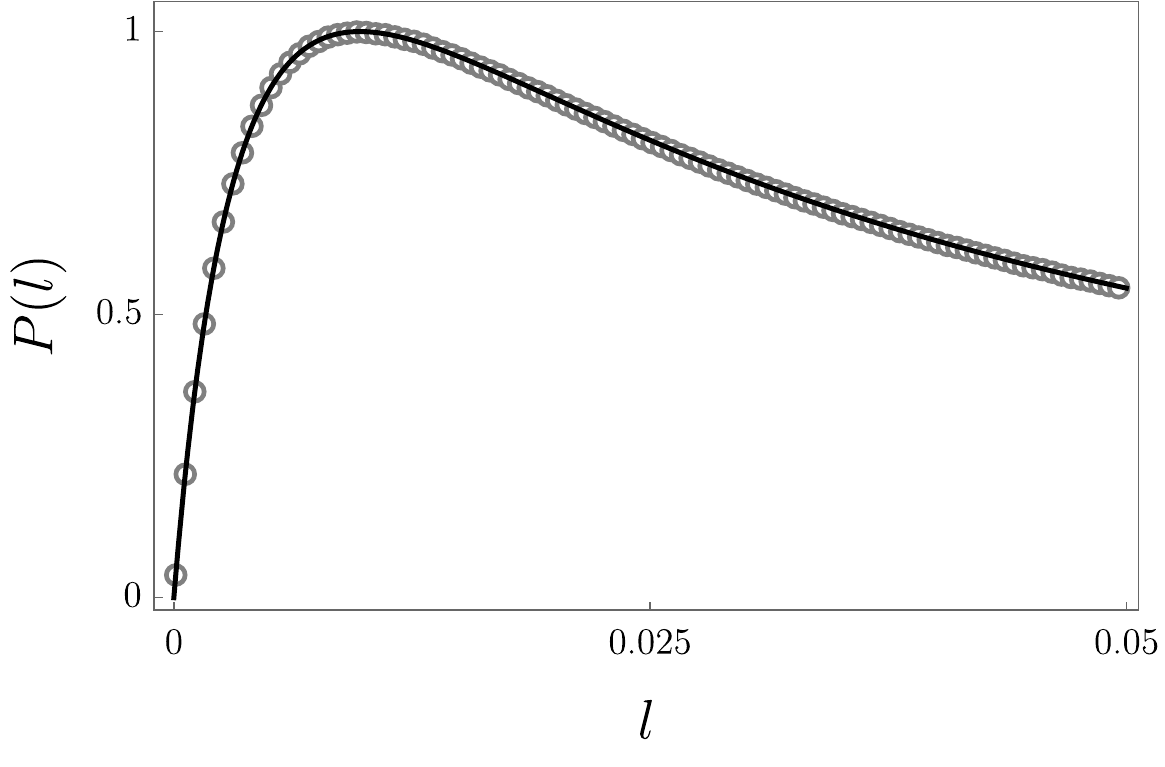}
    \caption{Success probability of the search algorithm on the hypercube of dimension 10 as a function of the weight $l$. Black curve is given by (\ref{ssl}), gray circles correspond to numerical simulation.}
    \label{fig:ssl}
\end{figure}

Turning to the run-time of the search algorithm, to rotate from $\ket{\alpha_2}$ to $-\ket{\alpha_2}$ the number of steps has to be taken approximately as
$$
T_1 \approx \frac{\pi}{\omega} \approx \frac{\pi\sqrt{2n}(l+n)}{(2n-1)\sqrt{l+\frac{n}{ 2^{n}}}}.
$$
For the optimal weight, we find
$$
T_1 \approx   \frac{n \pi (1+2^{-n})}{2n-1} 2^\frac{n}{2} \sim \frac{\pi}{2} 2^\frac{n}{2}  = O(\sqrt{N}) . 
$$

We can now utilize the derived results for search to perform state transfer between any two vertices using the switch approach \cite{skoupy:2022,Santos_2022}. We begin at the sender vertex $\vec{s}$ in the state $\ket{\vec{s},0}$. Note that at this stage the weighted loop is considered only on the sender vertex. The system is evolved for $T_1$ steps, reaching close to the equal-weight superposition state $\ket{\psi_0}$. Then we switch the marked vertex from the sender to the receiver $\vec{r}$, and evolve for $T_1$ more steps. With high fidelity the particle will be found in the loop at the receiver vertex, i.e. in the state $\ket{\vec{r},0}$. The total number of steps for state transfer with switch is then given by
\begin{equation}
\label{t:sta:switch} 
T_2 = 2 T_1 \approx  \frac{2n \pi (1+2^{-n})}{2n-1} 2^\frac{n}{2} \sim \pi 2^\frac{n}{2}.
\end{equation}


\section{State transfer between the antipodal vertices}
\label{sec:antipode} 

Let us consider the state transfer between the antipodal vertices (without loss of generality $\vec{0}$ and $\vec{1} = 1\ldots 1$). Due to the symmetry of this configuration the problem can be again reduced to a finite line of length $n$, with loops on both ends, i.e. in comparison with the search we have one additional basis state $\ket{n,\circlearrowleft} = \ket{\vec{1},0}$. The evolution operator is given by
$$
U'' = C''\cdot S'',
$$
where the shift reads
$$
S'' = S + \ketbra{0,\circlearrowleft}{0,\circlearrowleft} + \ketbra{n,\circlearrowleft}{n,\circlearrowleft},
$$
and the coin operator is decomposed as
$$
C'' = \left(\ketbra{0}{0} + \ketbra{n}{n}\right)\otimes C_0'  + \sum_{x=1}^{n-1} \ketbra{x}{x}\otimes C_x . 
$$

The initial state of the state transfer algorithm is the loop at the sender vertex $0$, i.e.
$$
\ket{\psi(0)} = \ket{0,\circlearrowleft}.
$$
We show that the algorithm evolves approximately in a five dimensional subspace spanned by $\{\ket{\psi_0},\ket{0,\circlearrowleft},\ket{n,\circlearrowleft},\ket{\psi_1},\ket{\psi_2}\}$, where we have denoted
$$
\ket{\psi_2} = \frac{1}{c} \sum_{x=0}^{n/2-2} \frac{1}{\sqrt{2{n-1\choose x}}} \left(\ket{n-x,L} - \ket{n-x-1,R}\right). 
$$
First, by direct calculation one can show that
$$
\ket{\beta_1} = \sqrt{\frac{l 2^n}{l 2^n +2n}}\ket{\psi_0} - \sqrt{\frac{n}{l 2^n +2n}}(\ket{0,\circlearrowleft} + \ket{n,\circlearrowleft}),
$$
is an exact eigenvector of $U''$ with eigenvalue 1. Next, consider the following four orthonormal states
\begin{align}
\nonumber \ket{\beta_2} & = \sqrt{\frac{2 n}{l 2^n + 2 n}}\ket{\psi_0} + \sqrt{\frac{l 2^{n-1}}{l 2^{n}+ 2n}}(\ket{0,\circlearrowleft} + \ket{n,\circlearrowleft}), \\
\nonumber    \ket{\beta_3} & = \frac{1}{\sqrt{2}} \left(\ket{\psi_1} - \ket{\psi_2}\right), \\
\nonumber    \ket{\beta_4} & = \frac{1}{\sqrt{2}}\left(\ket{0,\circlearrowleft} - \ket{n,\circlearrowleft}\right), \\
\label{basis:beta}     \ket{\beta_5} & = \frac{1}{\sqrt{2}} \left(\ket{\psi_1} + \ket{\psi_2}\right) .
\end{align}
The action of the evolution operator on these vectors is given by
\begin{align*}
    U'' \ket{\beta_2} = & \ket{\beta_2
    } - \frac{\sqrt{2 l(l+n \ 2^{1-n})}}{n+l}(\ket{0,\circlearrowleft} + \ket{n,\circlearrowleft}) - \frac{\sqrt{2 n(l+n \ 2^{1-n})}}{n+l}(\ket{1,L} + \ket{n-1,R}) , \\
    U'' \ket{\beta_3}  = &  \ket{\beta_3} - \frac{\sqrt{l n}}{c(n+l)} (\ket{0,\circlearrowleft} + \ket{n,\circlearrowleft} ) + \frac{l}{c(n+l)}(\ket{1,L} + \ket{n-1,R}) - \\
& - \frac{1}{c\sqrt{{n-1\choose \lfloor \frac{n}{2}\rfloor-2}}} \left( \frac{\lfloor\frac{n}{2}\rfloor-1}{n}(\ket{\lfloor \frac{n}{2}\rfloor-2,R} + \ket{\lfloor\frac{n+1}{2}\rfloor+2,L}) + \right. \\
& \left. +\frac{\sqrt{(\lfloor\frac{n}{2}\rfloor-1)(n-\lfloor\frac{n}{2}\rfloor+1)}}{n}(\ket{\lfloor\frac{n}{2}\rfloor,L} + \ket{\lfloor \frac{n+1}{2}\rfloor,R})\right), \\
    U'' \ket{\beta_4} = & \frac{n-l}{n+l}\ket{\beta_4} - \frac{\sqrt{2 l n}}{l+n} (\ket{1,L} - \ket{n-1,R}) , \\
U'' \ket{\beta_5}  = &  \ket{\beta_5}  - \frac{\sqrt{l n}}{c(n+l)} (\ket{0,\circlearrowleft} - \ket{n,\circlearrowleft} ) + \frac{l}{c(n + l)}(\ket{1,L} - \ket{n-1,R}) - \\
& - \frac{1}{c\sqrt{{n-1\choose \lfloor \frac{n}{2}\rfloor-2}}} \left( \frac{\lfloor\frac{n}{2}\rfloor-1}{n}(\ket{\lfloor \frac{n}{2}\rfloor-2,R} - \ket{\lfloor\frac{n+1}{2}\rfloor+2,L}) + \right. \\
& \left. +\frac{\sqrt{(\lfloor\frac{n}{2}\rfloor-1)(n-\lfloor\frac{n}{2}\rfloor+1)}}{n}(\ket{\lfloor\frac{n}{2}\rfloor,L} - \ket{\lfloor \frac{n+1}{2}\rfloor,R})\right) .
\end{align*}
From this we find that the diagonal matrix elements of $U''$ read
\begin{align*}
    \braketA{\beta_2}{U''}{\beta_2} = & 1-\frac{2 \left(l+ n\ 2^{1-n}\right)}{n+l} , \\
    \braketA{\beta_3}{U''}{\beta_3} = & \braketA{\beta_5}{U''}{\beta_5} = 1 -  \frac{l}{c^2(n+l)} - \frac{\lfloor\frac{n}{2}\rfloor-1}{c^2 n {n-1\choose \lfloor\frac{n}{2}\rfloor-2}}, \\
    \braketA{\beta_4}{U''}{\beta_4} = & 1 - \frac{2l}{n+l} .
\end{align*}
The only non-zero off-diagonal matrix elements are
\begin{align*}
    \braketA{\beta_3}{U''}{\beta_2} & = - \braketA{\beta_2}{U''}{\beta_3} = \frac{\sqrt{2n \left(l+ n 2^{1-n}\right)}}{c(n+l)} , \\
    \braketA{\beta_5}{U''}{\beta_4} & = - \braketA{\beta_4}{U''}{\beta_5} = \frac{\sqrt{2 l n}}{c(n+l)} .
\end{align*}
Hence, for large $n$ and small $l$ the evolution operator in the approximate invariant subspace is close to the effective unitary operator
$$
U''_{ef} = \begin{pmatrix}
1 & 0 & 0 & 0 & 0 \\
0 & \cos\omega_1 & -\sin\omega_1 & 0 & 0 \\
0 & \sin\omega_1 & \cos\omega_1 & 0 & 0 \\
0 & 0 & 0 & \cos\omega_2 & -\sin\omega_2 \\
0 & 0 & 0 & \sin\omega_2 & \cos\omega_2 
\end{pmatrix}, 
$$
where the angles $\omega_j$ are given by
\begin{align*}
\omega_1 & \approx  \frac{\sqrt{2n \left(l+ n 2^{1-n}\right)}}{c(n+l)} ,\quad \omega_2 \approx  \frac{\sqrt{2 l n}}{c(n+l)} . 
\end{align*}

To achieve state transfer with high fidelity we can tune the weight $l$ to make the frequencies harmonic. In particular, for 
\begin{equation}
\label{l:sta}
l = \frac{2}{3} \frac{n}{2^n},    
\end{equation}
we find (with the approximation (\ref{norm:approx}))
\begin{equation}
\label{sta:omega}
\omega_2 = \frac{\omega_1}{2} \approx \frac{2n-1}{\sqrt{3} n\left(1+\frac{2}{3}2^{-n}\right)} 2^{-\frac{n}{2}} \sim  \frac{2}{\sqrt{3}} 2^{-\frac{n}{2}}.    
\end{equation}
Note that for this value of $l$ the basis states $\ket{\beta_{1,2}}$ reduce to
\begin{align}
\nonumber  \ket{\beta_1} & = \frac{1}{2}\ket{\psi_0} - \sqrt{\frac{3}{8}}(\ket{0,\circlearrowleft} + \ket{n,\circlearrowleft}), \\
\label{beta:l}    \ket{\beta_2} & = \frac{\sqrt{3}}{2}\ket{\psi_0} + \sqrt{\frac{1}{8}}(\ket{0,\circlearrowleft} + \ket{n,\circlearrowleft}) .
\end{align}
The eigenvectors and eigenvalues of $U''_{ef}$ are then given by
\begin{align*}
    \ket{\omega_1^{(\pm)}} & = \frac{1}{\sqrt{2}} \left(\ket{\beta_2} \mp i\ket{\beta_3}\right), \quad \lambda_1^{(\pm)} = e^{\pm 2i \omega_2}, \\
        \ket{\omega_2^{(\pm)}} & = \frac{1}{\sqrt{2}} \left(\ket{\beta_4} \mp i\ket{\beta_5}\right), \quad \lambda_2^{(\pm)} = e^{\pm i \omega_2}.
\end{align*}
The initial and the final state (loops at the sender and the receiver vertex) of the state transfer can be decomposed as
\begin{align*}
    \ket{0,\circlearrowleft}  = & -\sqrt{\frac{3}{8}}\ket{\beta_1} + \frac{1}{\sqrt{8}} \ket{\beta_2} + \frac{1}{\sqrt{2}}\ket{\beta_4} \\
     = & -\sqrt{\frac{3}{8}}\ket{\beta_1} +\frac{1}{4} (\ket{\omega_1^{(+)}} + \ket{\omega_1^{(-)}}) + \frac{1}{2} (\ket{\omega_2^{(+)}} + \ket{\omega_2^{(-)}}), \\
     \ket{n,\circlearrowleft}  = &  -\sqrt{\frac{3}{8}}\ket{\beta_1} + \frac{1}{\sqrt{8}} \ket{\beta_2} - \frac{1}{\sqrt{2}}\ket{\beta_4} \\
     = & -\sqrt{\frac{3}{8}}\ket{\beta_1} +\frac{1}{4} (\ket{\omega_1^{(+)}} + \ket{\omega_1^{(-)}}) - \frac{1}{2} (\ket{\omega_2^{(+)}} + \ket{\omega_2^{(-)}}).
\end{align*}
The time evolution of the walk is then given by
\begin{align*}
\ket{\psi(t)} = &  -\sqrt{\frac{3}{8}}\ket{\beta_1} +\frac{1}{4} (e^{i 2\omega_2 t}\ket{\omega_1^{(+)}} + e^{- i 2\omega_2 t}\ket{\omega_1^{(-)}}) + \frac{1}{2} (e^{i \omega_2 t}\ket{\omega_2^{(+)}} + e^{- i \omega_2 t}\ket{\omega_2^{(-)}}) . \end{align*}
Hence, the fidelity of state transfer into the loop at the receiver vertex at time $t$ is equal to
\begin{equation}
\label{sta:fid:mostd}
   {\cal F}(t) = |\braket{n,\circlearrowleft}{\psi(t)}|^2 = \frac{1}{64}\left(3 + \cos\left(2\omega_2 t\right)  - 4\cos\left(\omega_2 t\right)\right)^2 , 
\end{equation}
which reaches unity for 
\begin{equation}
\label{sta:time}
    T_3 = \frac{\pi}{\omega_2} \approx \frac{\pi \sqrt{3} n\left(1+\frac{2}{3}2^{-n}\right)}{2n-1} 2^{-\frac{n}{2}} \sim \frac{\pi \sqrt{3}}{2}2^{\frac{n}{2}}.
\end{equation}
In comparison to state transfer with a switch (\ref{t:sta:switch}) the run-time is faster by a factor $\frac{\sqrt{3}}{2}$.

\begin{figure}
    \centering
    \includegraphics[width=0.65\textwidth]{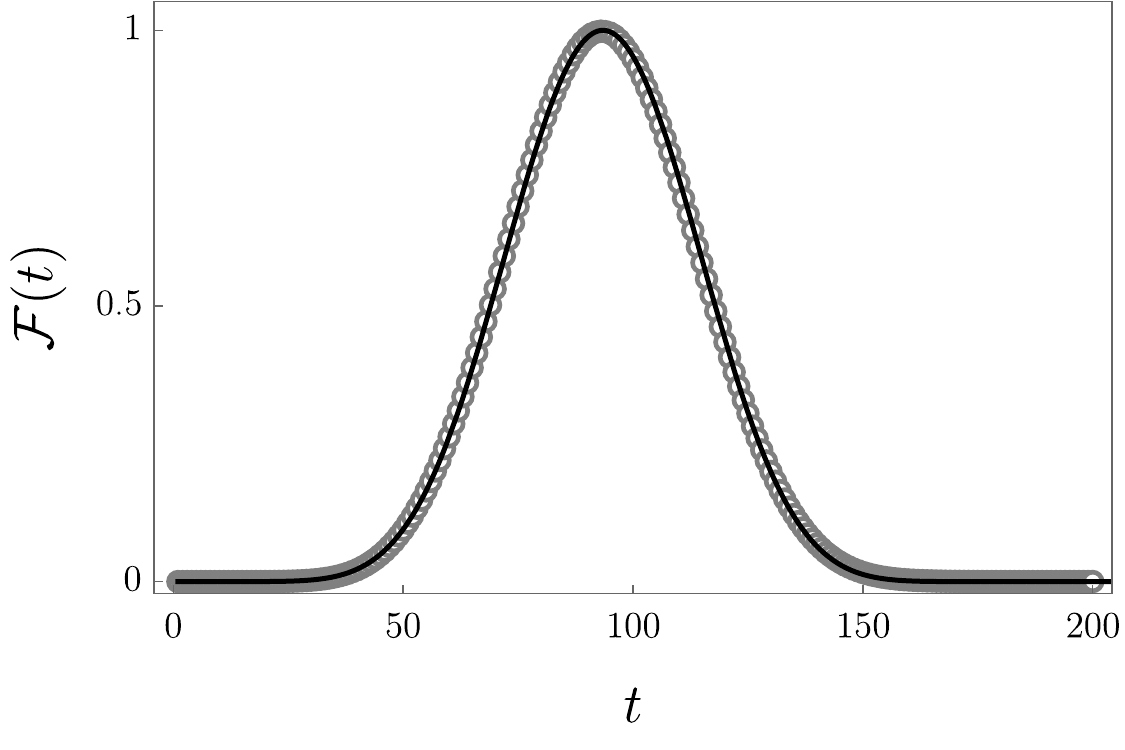}
    \caption{Fidelity of state transfer between the antipodal vertices on a hypercube of dimension 10. Black curve correspond to (\ref{sta:fid:mostd}) (with an exact value of $c$), gray circles are obtained from numerical simulation.}
    \label{fig:sta:mostd:fid}
\end{figure}

For illustration, we display in Figure \ref{fig:sta:mostd:fid} the fidelity of state transfer between the most distant vertices on a hypercube of dimension 10. 

We point out that the Grover walk on the hypercube without marked vertices is capable of state transfer to the antipodal vertex in polynomial time in $n$ \cite{Kempe2005}. However, this works only for the antipodes. As we illustrate with the numerical investigation in the following section, marking sender and receiver vertices with a weighted loop allows for state transfer at arbitrary distance.  

\section{State transfer between vertices of arbitrary distance}
\label{sec:arbitrary}

When the sender and the receiver vertices are not antipodes the reduction to a line is not possible anymore. For this reason, we turn to numerical simulations. Our investigations indicate that state transfer with high fidelity can be achieved for arbitrary distance $d$ with the same weight of the loops (\ref{l:sta}) and time of measurement (\ref{sta:time}) as for the state transfer to the antipodal vertex ($d=n$). Because of the symmetries of the hypercube it is sufficient to consider one particular vertex of a distance $d$ from the sender. Without loss of generality we consider the sender at the vertex $\vec{0}$ and the receiver at $\vec{r}$, where the Hamming weight of the $n$-bit string $\vec{r}$ is $d$. The initial and the target vectors of the state transfer algorithm are the loop at the sender and the receiver vertices, i.e. $\ket{\vec{0},0}$ and $\ket{\vec{r},0}$.

\begin{figure}[h]
    \centering
    \includegraphics[width=0.65\textwidth]{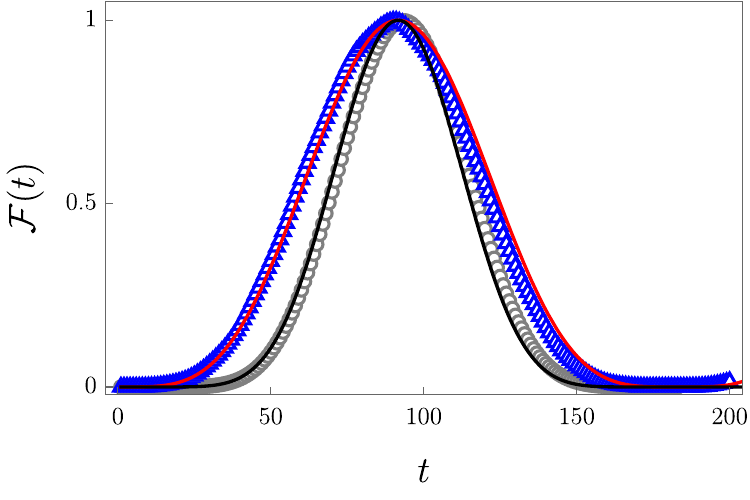}
    \caption{Fidelity of state transfer on a hypercube of dimension 10. We consider the receiver vertex at distance 5 (gray circles) and 1 (blue triangles) from the sender vertex. For $d=5$ the curve follows the result for the antipode (\ref{sta:fid:mostd}). For $d=1$ the peak of the fidelity is visibly wider and the course can be approximated by (\ref{fid:nn}).}
    \label{fig:fid:10}
\end{figure}

In Figure \ref{fig:fid:10} we display the course of fidelity of state transfer for a hypercube of dimension $n=10$ as a function of the number of steps $t$. We consider two choices of distance between the sender and the receiver vertex - $d=5$ depicted by the gray circles and $d=1$ corresponding to the blue triangles. For $d=5$, and in fact for all $d\neq 1$, the fidelity closely follows the curve (\ref{sta:fid:mostd}) derived for the state transfer to the antipode. However, in the case of state transfer between the direct neighbours the curve is less steep and the peak is visibly wider. Nevertheless, the maximal fidelity is reached at approximately the same time given by (\ref{sta:time}). The plot indicates that the algorithm behaves in a different way for $d=1$ and $d\geq 2$.

\begin{figure}
    \centering
    \includegraphics[width=0.65\textwidth]{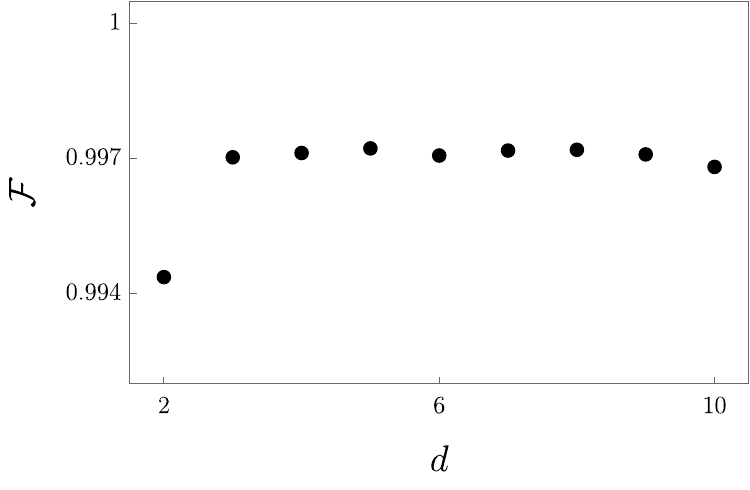}
    \caption{Fidelity of state transfer at time $T_3$ (\ref{sta:time}) in dependence on the distance between the sender and the receiver. We consider a hypercube of dimension 10.}
    \label{fig:sta:fid:d}
\end{figure}

\begin{figure}
    \centering
    \includegraphics[width=0.65\textwidth]{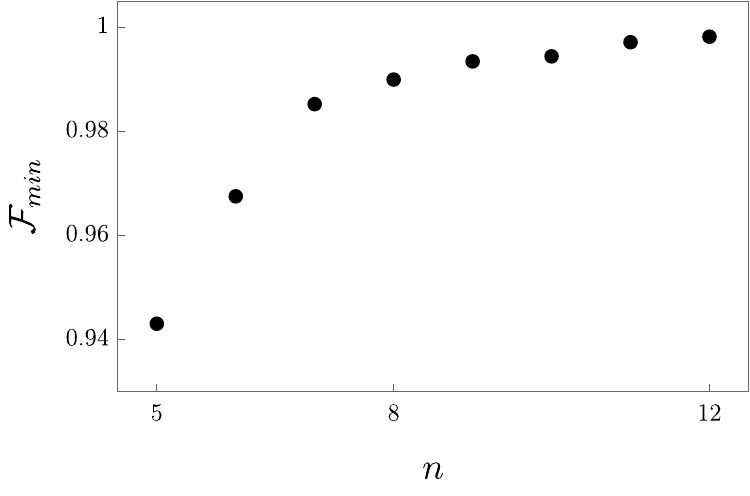}
    \caption{Minimal fidelity of state transfer at time $T_3$ (\ref{sta:time}) for a hypercube of dimension $n$.}
    \label{fig:min:fid:n}
\end{figure}

Let us first consider the case $d\geq 2$, where the evolution is close to the case of the state transfer to the antipode. In Figures \ref{fig:sta:fid:d} and \ref{fig:min:fid:n} we focus on the fidelity of state transfer. For Figure \ref{fig:sta:fid:d} we fix the dimension of the hypercube ($n=10$) and display the fidelity of state transfer in dependence on the distance $d$. Fidelity reaches values around 0.997 in all cases except for $d=2$, where it is close to 0.994. In Figure \ref{fig:min:fid:n} we consider the minimal fidelity for the state transfer on hypercubes of dimensions ranging from $n= 5$ to $n=12$. There is a clear increasing trend of the minimal fidelity.

\begin{figure}
    \centering
    \includegraphics[width=0.65\textwidth]{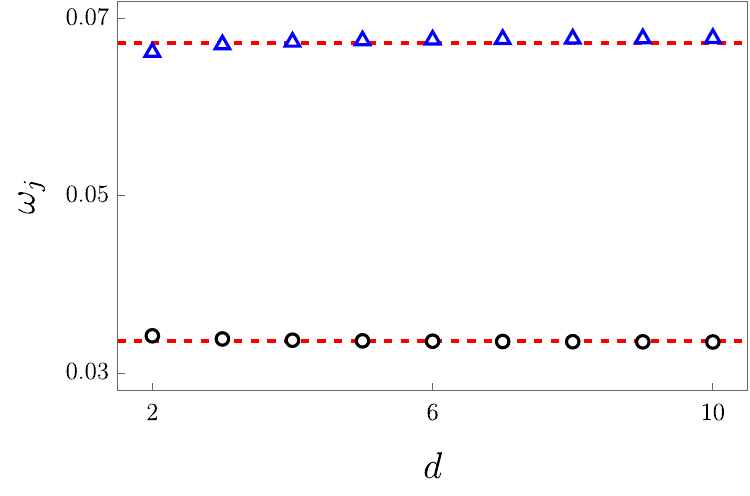}
    \includegraphics[width=0.65\textwidth]{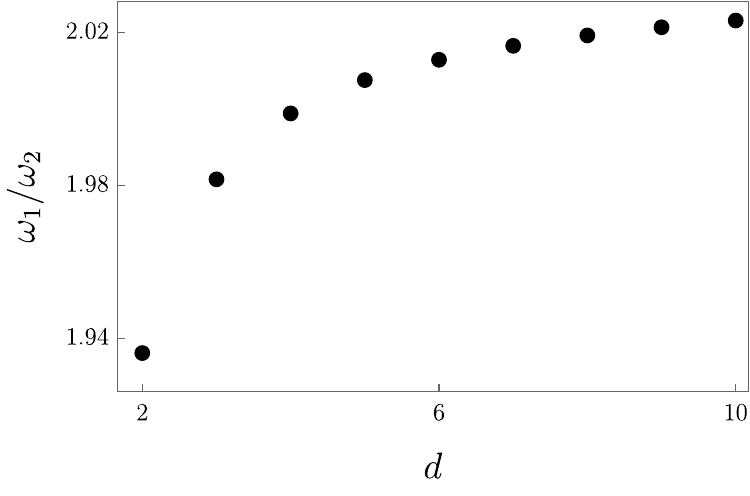}
    \caption{Upper plot shows numerically evaluated relevant eigenfrequencies for a hypercube of dimension $n=10$ depending on the distance between the sender and the receiver $d$. Circles correspond to the smaller angle $\omega_2$, triangles to $\omega_1$. Dashed lines highlight the values (\ref{sta:omega}) given from the state transfer to the antipode. The lower plot shows the ratio $\omega_1/\omega_2$ for different distances.}
    \label{fig:ratio}
\end{figure}

\begin{figure}
    \centering
    \includegraphics[width=0.65\textwidth]{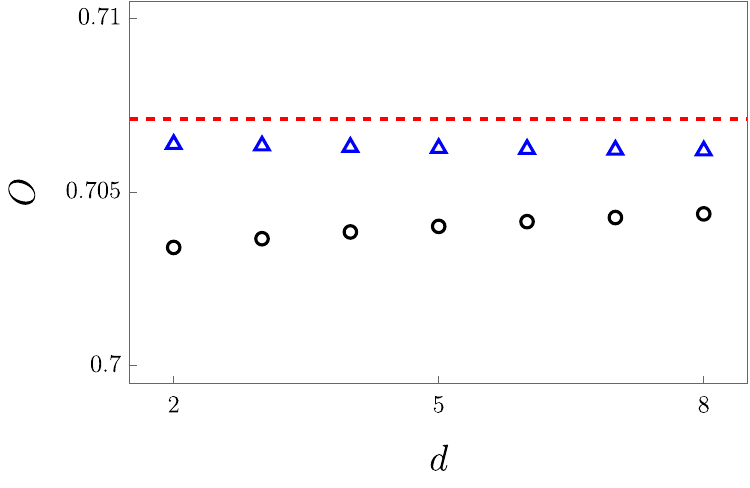}
    \caption{Overlap between the relevant eigenvectors of the evolution operator and the vectors $\ket{\gamma_2}$ and $\ket{\gamma_4}$, respectively. Black circles correspond to $\braket{\omega_1^{(\pm)}}{\gamma_2}$, blue triangles to $\braket{\omega_2^{(\pm)}}{\gamma_4}$. Dashed red line marks $1/\sqrt{2}$ which is the ideal theoretical value of the scalar products. Hypercube of dimension 8 is considered.}
    \label{fig:overlap}
\end{figure}

Let us now investigate the relevant eigenvalues and eigenvectors of the evolution operator. In analogy with (\ref{basis:beta}) and (\ref{beta:l}) we denote the following orthogonal vectors
\begin{eqnarray}
\label{gamma1} \ket{\gamma_1} & = & \frac{1}{2}\ket{\psi_0} - \sqrt{\frac{3}{8}}(\ket{\vec{0},0} + \ket{\vec{r},0}), \\
\nonumber \ket{\gamma_2} & = &\frac{\sqrt{3}}{2}\ket{\psi_0} + \sqrt{\frac{1}{8}}(\ket{\vec{0},0} + \ket{\vec{r},0}) , \\
\nonumber \ket{\gamma_4} & = & \frac{1}{\sqrt{2}}(\ket{\vec{0},0} - \ket{\vec{r},0}).
\end{eqnarray}
Direct calculation shows that $\ket{\gamma_1}$ is an exact eigenvector of $U''$ with eigenvalue 1 irrespective of $d$, i.e. including $d=1$. In fact, $\ket{\gamma_1}$ is an eigenstate of both the shift and the coin operator. To complete the approximate invariant subspace we turn to the numerical evaluation. Two pairs of complex conjugated eigenvalues $\lambda_j^{(\pm)} = e^{\pm i\omega_j}$, $j=1,2$ with the smallest non-zero eigenfrequencies $\omega_{j}$ are determined numerically, and the corresponding eigenvectors are found. The upper plot in Figure \ref{fig:ratio} shows the eigenfrequencies for the hypercube of dimension 10, confirming that for all distances $d=2,\ldots ,n$ they are close to the ideal theoretical values determined by (\ref{sta:omega}) for the state transfer to the antipodal vertex. The lower plot shows the ratio $\omega_1/\omega_2$. We see that the frequencies are close to a 2:1 resonance. In Figure \ref{fig:overlap} we consider the overlap of the corresponding numerically determined eigenvectors with the states $\ket{\gamma_2}$ and $\ket{\gamma_4}$ for a hypercube of dimension $n=8$. The scalar products are close to the ideal value $1/\sqrt{2}$ derived for the state transfer to the antipode. We conclude that the analytical results derived in Section \ref{sec:antipode} are well applicable to state transfer between vertices of arbitrary distance $d\geq 2$.

Turning to the case of sender and receiver being directly connected by an edge, we find that the distinction with the case of $d\geq 2$ is twofold. First, the projection of the initial and the target state onto $\ker(U''-1)$ is larger than what holds for an arbitrary distance (\ref{gamma1}). Let us denote by $d_s$ and $d_r$ the direction to the sender and receiver, respectively, so that e.g. $\ket{\vec{0},d_{r}}$ corresponds to the directed edge from the sender vertex $\vec{0}$ to the receiver vertex $\vec{r}$. A rather tedious calculation reveals that the projection is given by the following normalized eigenvector (note that $\ket{\vec{0},d_{r}}$ and $\ket{\vec{r},d_{s}}$ are not orthogonal to $\ket{\psi_0}$)
\begin{equation}
\ket{\gamma_0} = \frac{x}{\sqrt{2}} (\ket{\vec{0},0} + \ket{\vec{r},0}) + y \ket{\psi_0} +  \frac{z}{\sqrt{2}} (\ket{\vec{0},d_r} + \ket{\vec{r},d_s}),    
\end{equation}
where the coefficients $x,y,z$ read
\begin{align}
    \nonumber x & = \sqrt{\frac{3a}{b}}, \quad a = 2^n + 2n -4, \quad b = 3\cdot 2^n + 8n -12 -2^{2-n}, \\
    \nonumber  y & = -\frac{2(n-1)}{\sqrt{a b}}, \quad z = -\frac{\sqrt{n 2^{n+1}}(1-2^{1-n})}{\sqrt{a b}} .
\end{align}
For large $n$, we find that $x$ approaches unity while $y$ and $z$ vanish according to
\begin{align*}
\nonumber x = 1- O(2^{-n}), \quad y = O(n 2^{-n}), \quad z = O(\sqrt{n 2^{-n}}) .
\end{align*}
Therefore, for large $n$ the vector
$$
\ket{\tilde{\gamma_0}} = \frac{1}{\sqrt{2}}(\ket{\vec{0},0} + \ket{\vec{r},0}),
$$
is an approximate eigenvector corresponding to 1. 

\begin{figure}
    \centering
    \includegraphics[width=0.65\textwidth]{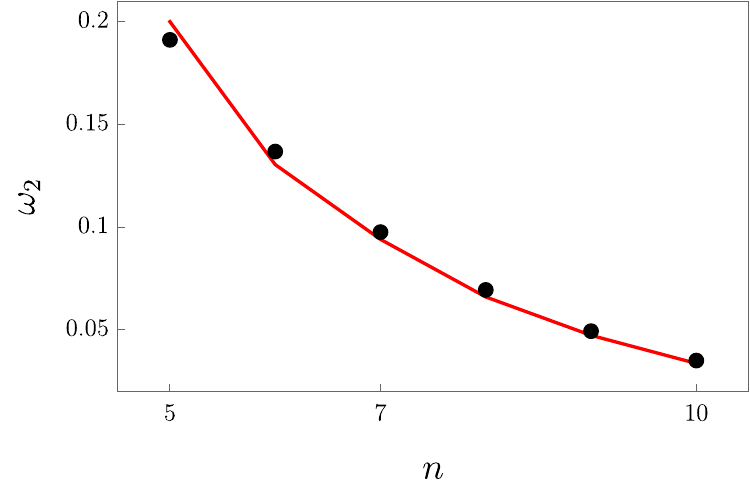}
    \caption{Comparison of numerically evaluated relevant eigenvalues $\lambda_2^{(\pm)}$ with the analytical value  for the case of state transfer between the neighbouring vertices. Black dots correspond to the phase of $\lambda_2^{(+)}$ for different dimensions of the hypercube $n$. The red curve is given by  (\ref{sta:omega}).}
    \label{fig:nn:omega}
\end{figure}

\begin{figure}
    \centering
    \includegraphics[width=0.65\textwidth]{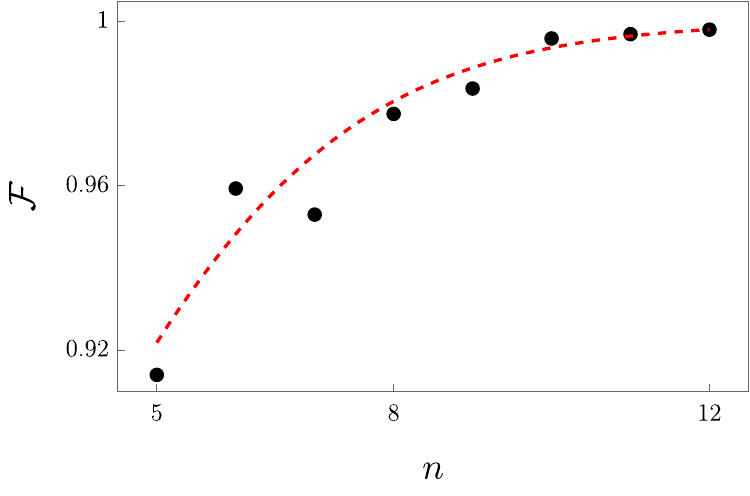}
    \caption{Fidelity of state transfer to a neighbouring vertex at time $T_3$ for a given dimension of the hypercube $n$. Black dots are obtained from numerical simulations. The red curve is taken from the approximation (\ref{fmax:nn}).}
    \label{fig:nn:f}
\end{figure}

Second distinction is in the relevant part of the spectrum of the evolution operator. Indeed, numerical evaluation indicates that the pair $\lambda_1^{(\pm)}$ is no longer present in the spectrum of $U''$. On the other, for the remaining pair $\lambda_2^{(\pm)}$ the angle $\omega_2$ is close to the formula (\ref{sta:omega}), see Figure \ref{fig:nn:omega}. Hence, the approximate invariant subspace is only three-dimensional, consisting of $\ket{\gamma_0}$ and $\ket{\omega_2^{(\pm)}}$. The initial and the target vectors of the state transfer algorithm are decomposed into this basis as
\begin{align*}
    \ket{\vec{0},0}  = & \frac{x}{\sqrt{2}}\ket{\gamma_0} + \frac{1}{2} (\ket{\omega_2^{(+)}} + \ket{\omega_2^{(-)}}) +\ket{\delta_s}, \\
     \ket{\vec{r},0}  = &  \frac{x}{\sqrt{2}}\ket{\gamma_0} - \frac{1}{2} (\ket{\omega_2^{(+)}} + \ket{\omega_2^{(-)}}) + \ket{\delta_r},
\end{align*}
where $\ket{\delta_{r,s}}$ are some residual non-normalized vectors, which vanish for large $n$ as $x$ tends to 1. The state of the system evolves according to
$$
\ket{\psi(t)} = \frac{x}{\sqrt{2}}\ket{\gamma_0} +\frac{1}{2}(e^{-i\omega_2 t}\ket{\omega_2^{(+)}} + e^{i\omega_2 t}\ket{\omega_2^{(-)}}) + \ket{\delta_s(t)}.
$$
Neglecting the residual vectors, the fidelity of state transfer at time $t$ then follows the curve
\begin{equation}
\label{fid:nn}
   {\cal F}(t) = |\braket{\vec{r},0}{\psi(t)}|^2 = \frac{1}{4}(x^2-\cos(\omega_2 t))^2.
\end{equation}
At time $T_3$ (\ref{sta:time}) fidelity reaches the value
\begin{equation}
\label{fmax:nn}
{\cal F} = \frac{1}{4}(1+x^2)^2.
\end{equation}
In Figure \ref{fig:nn:f} we compare this result with the fidelity at time $T_3$ obtained from numerical simulation. We see that the fidelity approaches unity with increasing dimension of the hypercube $n$. For large $n$, where we can replace $\ket{\gamma_0}$ with $\ket{\tilde{\gamma_0}}$ and $x$ with 1, we find that the fidelity at time $t$ reads
$$
{\cal F}(t) = \sin^4{\left(\frac{\omega_2 t}{2}\right)}, 
$$
which reaches unity for number of steps equal to $T_3$. This explains the broader peak in Figure \ref{fig:fid:10} in comparison with the case $d=5$.

\section{Conclusions}
\label{sec:concl}

Two approaches to state transfer on a hypercube were investigated in detail. The first one utilizes quantum walk search, where we evolve from the loop at the sender vertex to the initial state of the search algorithm, switch the marked vertex from the sender to the receiver, and evolve to the loop at the receiver vertex. It was shown that high fidelity of state transfer between any pair of vertices is achieved for the weight given by the ratio of the degree of the vertex to the total number of vertices. The run-time of the state transfer with a switch is then twice the number of steps of the search. In the second approach we have considered marking both the sender and the receiver vertices simultaneously. The case of antipodal vertices was investigated analytically, and we have shown that the optimal weight is 2/3 of the weight for the search. This results in a run-time faster by a factor of $\sqrt{3}/2$ when compared to the state transfer with a switch. Utilizing numerical simulations we have illustrated that the second approach works well for arbitrary distance between the sender and the receiver, although the evolution differs for the case of direct neighbours. This confirms that on a hypercube network any two parties can establish communication without the need to know each others position and using only local operations and global properties of the network (dimension of the hypercube $n$). In general, the fidelity of state transfer improves with increasing dimension $n$.

When the sender and the receiver are marked simultaneously the evolution is determined by two pairs of complex conjugated eigenvalues. State transfer with high fidelity was achieved by tuning the weight of the loops such that the two relevant frequencies became harmonic. It is an open question if such an effect can be demonstrated on different graphs. In addition, the ratio of the optimal weights for the state transfer and search, was shown to be 2/3 for the hypercube. One can ask if this is specific for a hypercube or it applies to a broader set of graphs.

Finally, let us briefly comment on the experimental implementations of quantum walk state transfer. Most of the experiments realized state transfer over a linear graph structure. Photonic chips \cite{grafe_integrated_2016} with engineered coupled waveguides are ideal for the implementations of continuous time quantum walk protocols. State transfer of a photon over a line was realized in \cite{perez-leija_coherent_2013}, and a more recent experiment performed a state transfer of a photon from an entangled pair \cite{chapman_experimental_2016}. Another experiment utilized superconducting qubit chain \cite{li_qubit_chain_2018}. Going beyond linear structures, a photonic chip has implemented hexagonal graphs \cite{tang_experimental_2018}. Concerning the discrete-time case, a photonic time-multiplexing set-up \cite{nitsche_quantum_2016} was utilized for realization of a quantum routing protocol \cite{zhan_perfect_2014} for a line. A similar protocol \cite{shang2018} was recently implemented on a quantum computer \cite{shang_experimental_2020}, performing a state transfer on a complete graph with 2, 4 or 8 vertices. The hypercube graph investigated in the present paper is rather complicated to be implemented physically on a photonic chip or encoded in time-multiplexing scenario, except for small dimensions $n=2,3$. However, one can consider implementation of state transfer algorithm on such graphs on a universal quantum computer or in a dedicated programmable qubit array \cite{gong_quantum_2021}. 

\ack

Both authors are grateful for financial support from RVO 14000 and ”Centre for Advanced Applied Sciences”, Registry No. CZ.02.1.01/0.0/0.0/16 019/0000778, supported by the Operational Programme Research, Development and Education, co-financed by the European Structural and Investment Funds. M\v S acknowledges the financial support from Czech Grant Agency project number GA\v CR 23-07169S. SS is grateful for financial support from SGS22/181/OHK4/3T/14.

We would like to express our gratitude to prof. Igor Jex for stimulating discussions and support over the years. Live long and prosper!

\section*{References}

\bibliography{biblio}

\begin{thebibliography}{10}

\bibitem{Aharonov1993}
Y.~Aharonov, L.~Davidovich, and N.~Zagury.
\newblock Quantum random walks.
\newblock {\em Phys. Rev. A}, 48:1687--1690, 1993.

\bibitem{Meyer1996}
D.~A. Meyer.
\newblock From quantum cellular automata to quantum lattice gases.
\newblock {\em J. Stat. Phys.}, 85:551--574, 1996.

\bibitem{Farhi1998}
E~Farhi and S~Gutmann.
\newblock Quantum computation and decision trees.
\newblock {\em Phys. Rev. A}, 58:915--928, Aug 1998.

\bibitem{bose2003}
S.~Bose.
\newblock Quantum communication through an unmodulated spin chain.
\newblock {\em Phys. Rev. Lett.}, 91:207901, 2003.

\bibitem{christandl_perfect_2004}
M.~Christandl, N.~Datta, A.~Ekert, and A.~J. Landahl.
\newblock Perfect {State} {Transfer} in {Quantum} {Spin} {Networks}.
\newblock {\em Phys. Rev. Lett.}, 92:187902, 2004.

\bibitem{christandl_perfect_2005}
M.~Christandl, N.~Datta, T.~C. Dorlas, A.~Ekert, A.~Kay, and A.~J. Landahl.
\newblock Perfect transfer of arbitrary states in quantum spin networks.
\newblock {\em Phys. Rev. A}, 71:032312, 2005.

\bibitem{plenio_high_2005}
M.~B. Plenio and F.~L. Semião.
\newblock High efficiency transfer of quantum information and multiparticle
  entanglement generation in translation-invariant quantum chains.
\newblock {\em New J. Phys.}, 7:73, 2005.

\bibitem{bose_2007}
S.~Bose.
\newblock Quantum communication through spin chain dynamics: an introductory
  overview.
\newblock {\em Contemp. Phys.}, 48:13--30, 2007.

\bibitem{gualdi_perfect_2008}
G.~Gualdi, V.~Kostak, I.~Marzoli, and P.~Tombesi.
\newblock Perfect state transfer in long-range interacting spin chains.
\newblock {\em Phys. Rev. A}, 78:022325, 2008.

\bibitem{kay_perfect_2010}
A.~Kay.
\newblock {Perfect}, {efficient}, {state} {transfer} {and} {its} {application}
  {as} {a} {constructive} {tool}.
\newblock {\em Int. J. Quantum Inf.}, 8:641--676, 2010.

\bibitem{kostak_perfect_2007}
V.~Kostak, G.~M. Nikolopoulos, and I.~Jex.
\newblock Perfect state transfer in networks of arbitrary topology and coupling
  configuration.
\newblock {\em Phys. Rev. A}, 75:042319, 2007.

\bibitem{nikolopoulos_analysis_2012}
G.~M. Nikolopoulos, A.~Hoskovec, and I.~Jex.
\newblock Analysis and minimization of bending losses in discrete quantum
  networks.
\newblock {\em Phys. Rev. A}, 85:062319, 2012.

\bibitem{hoskovec_decoupling_2014}
A.~Hoskovec, H.~Frydrych, I.~Jex, and G.~Alber.
\newblock Decoupling {Bent} {Quantum} {Networks}.
\newblock In {\em 2014 {International} {Symposium} on {Information} {Theory}
  and {Its} {Applications} (isita)}, pages 172--175, New York, 2014. IEEE.

\bibitem{frydrych_selective_2015}
H.~Frydrych, A.~Hoskovec, I.~Jex, and G.~Alber.
\newblock Selective dynamical decoupling for quantum state transfer.
\newblock {\em J. Phys. B - At. Mol. Opt. Phys.}, 48:025501, 2015.

\bibitem{hoskovec_dynamical_2022}
A.~Hoskovec and I.~Jex.
\newblock Dynamical decoupling and {NNN} discrete quantum networks.
\newblock {\em Int. J. Quantum Inf.}, 20:2250009, 2022.

\bibitem{kendon2011}
V.~M. Kendon and C.~Tamon.
\newblock {Perfect state transfer in quantum walks on graphs}.
\newblock {\em {J. Comput. Theor. Nanosci.}}, {8}:{422}, {2011}.

\bibitem{godsil_state_2012}
Ch. Godsil.
\newblock State transfer on graphs.
\newblock {\em Discrete Math.}, 312:129--147, 2012.

\bibitem{godsil_state_2020}
Ch. Godsil, K.~Guo, M.~Kempton, G.~Lippner, and F.~Münch.
\newblock State transfer in strongly regular graphs with an edge perturbation.
\newblock {\em J. Comb. Theory, Series A}, 172:105181, 2020.

\bibitem{coutinho_perfect_2015}
G.~Coutinho, C.~Godsil, K.~Guo, and F.~Vanhove.
\newblock Perfect state transfer on distance-regular graphs and association
  schemes.
\newblock {\em Linear Algebra Appl.}, 478:108--130, 2015.

\bibitem{chen_pair_2020}
Q.~Chen and Ch. Godsil.
\newblock Pair state transfer.
\newblock {\em Quantum Inf. Process.}, 19:321, 2020.

\bibitem{kurzynski2011}
P.~Kurzynski and A.~Wojcik.
\newblock {Discrete-time quantum walk approach to state transfer}.
\newblock {\em {Phys. Rev. A}}, {83}:{062315}, {2011}.

\bibitem{yalcinkaya2015}
I.~Yalcinkaya and Z.~Gedik.
\newblock {Qubit state transfer via discrete-time quantum walks}.
\newblock {\em {J. Phys. A}}, {48}:{225302}, {2015}.

\bibitem{shang2018}
Y.~Shang, Y.~Wang, M.~Li, and R.~Q. Lu.
\newblock Quantum communication protocols by quantum walks with two coins.
\newblock {\em EPL}, 124:60009, 2018.

\bibitem{chen2019}
X.~B. Chen, Y.~L. Wang, G.~Xu, and Y.~X. Yang.
\newblock Quantum network communication with a novel discrete-time quantum
  walk.
\newblock {\em IEEE ACCESS}, 7:13634, 2019.

\bibitem{zhan_quantum_2021}
H.~Zhan.
\newblock Quantum walks on embeddings.
\newblock {\em J. Algebr. Comb.}, 53:1187--1213, 2021.

\bibitem{kubota_perfect_2022}
S.~Kubota and E.~Segawa.
\newblock Perfect state transfer in {Grover} walks between states associated to
  vertices of a graph.
\newblock {\em Linear Algebra Appl.}, 646:238--251, 2022.

\bibitem{chan:2021}
A.~Chan and H.~Zhan.
\newblock Pretty good state transfer in discrete-time quantum walks.
\newblock {\em arXiv:2105.03762}.

\bibitem{guo:2022}
K.~Guo and V.~Schmeits.
\newblock Perfect state transfer in quantum walks on orientable maps.
\newblock {\em arXiv:2211.12841}.

\bibitem{aaronson_quantum_2003}
S.~Aaronson and A.~Ambainis.
\newblock Quantum search of spatial regions.
\newblock In {\em 44th {Annual} {IEEE} {Symposium} on {Foundations} of
  {Computer} {Science}, {Proceedings}}, pages 200--209. Ieee Computer Soc,
  2003.

\bibitem{Shenvi2003}
N.~Shenvi, J.~Kempe, and K.~B. Whaley.
\newblock Quantum random-walk search algorithm.
\newblock {\em Phys. Rev. A}, 67:052307, 2003.

\bibitem{Potocek2009}
V.~Poto{\v{c}}ek, A.~G{\'a}bris, T.~Kiss, and I.~Jex.
\newblock Optimized quantum random-walk search algorithms on the hypercube.
\newblock {\em Phys. Rev. A}, 79:012325, 2009.

\bibitem{hein2009:search}
B.~Hein and G.~Tanner.
\newblock {Quantum search algorithms on the hypercube}.
\newblock {\em {J. Phys. A}}, {42}:{085303}, {2009}.

\bibitem{Childs2004}
A.~M. Childs and J.~Goldstone.
\newblock Spatial search by quantum walk.
\newblock {\em Phys. Rev. A}, 70:022314, 2004.

\bibitem{childs_2004b}
A.~M. Childs and J.~Goldstone.
\newblock Spatial search and the dirac equation.
\newblock {\em Phys. Rev. A}, 70:042312, 2004.

\bibitem{Ambainis2005}
A.~Ambainis, J.~Kempe, and A.~Rivosh.
\newblock Coins make quantum walks faster.
\newblock In {\em Proc.~16th ACM-SIAM symposium on Discrete algorithms}, pages
  1099--1108. Society for Industrial and Applied Mathematics, 2005.

\bibitem{reitzner2009}
D.~Reitzner, M.~Hillery, E.~Feldman, and V.~Bužek.
\newblock Quantum searches on highly symmetric graphs.
\newblock {\em Phys. Rev. A}, 79:012323, 2009.

\bibitem{chakraborty2016}
S.~Chakraborty, L.~Novo, A.~Ambainis, and Y.~Omar.
\newblock Spatial search by quantum walk is optimal for almost all graphs.
\newblock {\em Phys. Rev. Lett.}, 116:100501, 2016.

\bibitem{ambainis_quadratic_2020}
A.~Ambainis, A.~Gilyen, S.~Jeffery, and M.~Kokainis.
\newblock Quadratic {Speedup} for {Finding} {Marked} {Vertices} by {Quantum}
  {Walks}.
\newblock In {\em Proceedings of the 52nd {Annual} {Acm} {Sigact} {Symposium}
  on {Theory} of {Computing} (stoc '20)}, pages 412--424. Assoc Computing
  Machinery, 2020.

\bibitem{apers_quadratic_2022}
S.~Apers, S.~Chakraborty, L.~Novo, and J.~Roland.
\newblock Quadratic {Speedup} for {Spatial} {Search} by {Continuous}-{Time}
  {Quantum} {Walk}.
\newblock {\em Phys. Rev. Lett.}, 129:160502, 2022.

\bibitem{hein2009}
B.~Hein and G.~Tanner.
\newblock Wave communication across regular lattice.
\newblock {\em Phys. Rev. Lett.}, 103:260501, 2009.

\bibitem{barr2014}
K.~Barr, T.~Proctor, D.~Allen, and V.~M. Kendon.
\newblock {Periodicity and perfect state transfer in quantum walks on variants
  of cycles}.
\newblock {\em {Quantum Inf. Comput.}}, {14}:{417}, {2014}.

\bibitem{stefanak2016}
M.~Štefaňák and S.~Skoupý.
\newblock Perfect state transfer by means of discrete-time quantum walk on
  highly symmetric graphs.
\newblock {\em Phys. Rev. A}, 94:022301, 2016.

\bibitem{stefanak2017}
M.~Štefaňák and S.~Skoupý.
\newblock Perfect state transfer by means of discrete-time quantum walk on
  complete bipartite graphs.
\newblock {\em Quantum Inf. Process.}, 16:72, 2017.

\bibitem{zhan2019}
H.~Zhan.
\newblock {An infinite family of circulant graphs with perfect state transfer
  in discrete quantum walks}.
\newblock {\em {Quantum Inf. Process.}}, {18}:{369}, {2019}.

\bibitem{cao2019}
W.~F. Cao, Y.~G. Yang, D.~Li, J.~R. Dong, Y.~H. Zhou, and W.~M. Shi.
\newblock Quantum state transfer on unsymmetrical graphs via discrete-time
  quantum walk.
\newblock {\em Mod. Phys. Lett. A}, 34:1950317, 2019.

\bibitem{skoupy:2022}
S.~Skoup\'y and M.~\ifmmode \check{S}\else
  \v{S}\fi{}tefa\ifmmode~\check{n}\else \v{n}\fi{}\'ak.
\newblock Quantum-walk-based state-transfer algorithms on the complete
  $m$-partite graph.
\newblock {\em Phys. Rev. A}, 103:042222, 2021.

\bibitem{Santos_2022}
R.~A.~M. Santos.
\newblock Quantum state transfer on the complete bipartite graph.
\newblock {\em J. Phys. A: Math. Theor.}, 55:125301, 2022.

\bibitem{wong2015}
T.~G. Wong.
\newblock Grover search with lackadaisical quantum walks.
\newblock {\em J. Phys. A}, 48:435304, 2015.

\bibitem{wong2018}
T.~G. Wong.
\newblock {Faster search by lackadaisical quantum walk}.
\newblock {\em {Quantum Inf. Process.}}, {17}:{68}, {2018}.

\bibitem{rhodes2019}
M.~L. Rhodes and T.~G. Wong.
\newblock {Search by lackadaisical quantum walks with nonhomogeneous weights}.
\newblock {\em {Phys. Rev. A}}, {100}:{042303}, {2019}.

\bibitem{rhodes2020}
M.~L. Rhodes and T.~G. Wong.
\newblock {Search on vertex-transitive graphs by lackadaisical quantum walk}.
\newblock {\em {Quantum Inf. Process.}}, {19}:{334}, {2020}.

\bibitem{chiang2020}
C.~Chiang.
\newblock Overview: recent development and applications of reduction and
  lackadaisicalness techniques for spatial search quantum walk in the near
  term.
\newblock {\em Quantum Inf. Process.}, 19:364, 2020.

\bibitem{hoyer2020}
P.~H\o{}yer and Z.~Yu.
\newblock {Analysis of Lackadaisical Quantum Walks}.
\newblock {\em Quant. Inf. Comput.}, 20:14, 2020.

\bibitem{Kempe2005}
J.~Kempe.
\newblock Discrete quantum walks hit exponentially faster.
\newblock {\em Prob. Theor. Rel. Fields}, 133:215--235, 2005.

\bibitem{grafe_integrated_2016}
M.~Gräfe, R.~Heilmann, M.~Lebugle, D.~Guzman-Silva, A.~Perez-Leija, and
  A.~Szameit.
\newblock Integrated photonic quantum walks.
\newblock {\em J. Opt.}, 18:103002, 2016.

\bibitem{perez-leija_coherent_2013}
A.~Perez-Leija~et al.
\newblock Coherent quantum transport in photonic lattices.
\newblock {\em Phys. Rev. A}, 87:012309, 2013.

\bibitem{chapman_experimental_2016}
R.~J. Chapman~et al.
\newblock Experimental perfect state transfer of an entangled photonic qubit.
\newblock {\em Nat. Commun.}, 7:11339, 2016.

\bibitem{li_qubit_chain_2018}
X.~Li~et al.
\newblock Perfect quantum state transfer in a superconducting qubit chain with
  parametrically tunable couplings.
\newblock {\em Phys. Rev. Appl.}, 10:054009, 2018.

\bibitem{tang_experimental_2018}
H.~Tang~et al.
\newblock Experimental quantum fast hitting on hexagonal graphs.
\newblock {\em Nat. Photonics}, 12:754, 2018.

\bibitem{nitsche_quantum_2016}
T.~Nitsche, F.~Elster, J.~Novotný, A.~Gábris, I.~Jex, S.~Barkhofen, and Ch.
  Silberhorn.
\newblock Quantum walks with dynamical control: graph engineering, initial
  state preparation and state transfer.
\newblock {\em New J. Phys.}, 18:063017, 2016.

\bibitem{zhan_perfect_2014}
X.~Zhan, H.~Qin, Z.~Bian, J.~Li, and P.~Xue.
\newblock Perfect state transfer and efficient quantum routing: {A}
  discrete-time quantum-walk approach.
\newblock {\em Phys. Rev. A}, 90:012331, 2014.

\bibitem{shang_experimental_2020}
Y.~Shang and M.~Li.
\newblock Experimental realization of state transfer by quantum walks with two
  coins.
\newblock {\em Quantum Sci. Technol.}, 5:015005, 2020.

\bibitem{gong_quantum_2021}
M.~Gong~et al.
\newblock Quantum walks on a programmable two-dimensional 62-qubit
  superconducting processor.
\newblock {\em Science}, 372:948, 2021.

\end{thebibliography}
\bibliographystyle{unsrt}

\end{document}